\documentclass[prc,twocolumn,aps,showpacs,floatfix,nofootinbib,letterpaper,preprintnumbers]{revtex4-1}
\usepackage{epsfig}
\usepackage{amssymb}
\usepackage{amsmath}
\usepackage{amsfonts}
\usepackage{color,graphicx}
\usepackage{bm}
\usepackage{hyperref}

\newcommand{\cg}[6]{(#1 \, #2 \,\, #3 \, #4|#5 \, #6)}

\definecolor{DarkBlue}{RGB}{10,10,190}
\hypersetup{
    pdftoolbar=true,        
    pdfmenubar=true,        
    pdffitwindow=false,     
    pdfstartview={FitH},    
    pdftitle={},    
    pdfauthor={},     
    pdfsubject={},   
    pdfnewwindow=true,      
    colorlinks=true,        
    allcolors=DarkBlue
}

\begin{document}

\preprint{INT-PUB-17-040}

\title{Nuclear deformation in the laboratory frame}
\author{C.N.~Gilbreth,$^{1}$ Y. Alhassid,$^{2}$ and G.F.~Bertsch$^{1,3}$ }
\affiliation{$^{1}$Institute of Nuclear Theory,
  Box 351550, University of Washington, Seattle, WA 98915\\
  $^{2}$Center for Theoretical Physics, Sloane Physics
  Laboratory, Yale University, New Haven, CT 06520\\
  $^{3}$Department of Physics, Box 351560, University of Washington, Seattle, WA 98195
}
\date{\today}
\def\lb{\langle}
\def\rb{\rangle}
\def\ni{\noindent}
\def\be{\begin{equation}}
\def\ee{\end{equation}}
\def\sumk{\sum_k}
\def\ad{a^\dagger_k}
\def\adb{a^\dagger_{\bar k}}
\def\a{a_k}
\def\ab{a_{\bar k}}
\def\Tr{{\rm Tr}}
\def\tr{{\rm tr}}
\def\Re{{\rm Re\,}}

\begin{abstract}
We develop a formalism for calculating the distribution of the axial quadrupole operator in the laboratory frame  within the rotationally invariant framework of the configuration-interaction shell model.  The calculation is carried out using a finite-temperature auxiliary-field quantum Monte Carlo method.  We apply this formalism to isotope chains of even-mass samarium and neodymium nuclei, and show that the quadrupole distribution provides a model-independent signature of nuclear deformation.  Two technical advances are described that greatly facilitate the calculations.  The first is to exploit the rotational invariance of the underlying Hamiltonian to reduce the statistical fluctuations in the Monte Carlo calculations. The second is to determine quadruple invariants from the distribution of the axial quadrupole operator. This allows us to extract effective values of the intrinsic quadrupole shape parameters without invoking an intrinsic frame or a mean-field approximation. 
\end{abstract}

\pacs{21.60.Cs,  21.60.Ka, 21.10.Ma,  02.70.Ss}

\maketitle

\section{Introduction}   

Although the Hamiltonian of an atomic nucleus is rotationally invariant, open-shell nuclei often exhibit features that are qualitatively well-described by simple models in which the nucleus is deformed rather than spherical. In particular, deformed nuclei show rotational bands of energy levels consistent with the model of an axially symmetric rigid rotor~\cite{BM75}. 

Deformation is typically introduced through a mean-field approximation that breaks the rotational symmetry of the underlying Hamiltonian. For instance, the axial quadrupole operator $\hat{Q}_{2\mu}$, which measures quadrupole deformations, acquires a nonzero expectation in the solution to the Hartree-Fock (HF) or Hartree-Fock-Bogoliubov (HFB) equations for deformed nuclei.
However, the exact expectation values of $\hat{Q}_{2\mu}$ are always zero for a nucleus described by a rotationally-invariant Hamiltonian. Because atomic nuclei do exhibit signatures of deformation, yet they are described by a rotationally invariant Hamiltonian, it is of interest to be able to extract information about nuclear deformation within a framework that preserves rotational invariance and without invoking a mean-field approximation.   

In Ref.~\cite{al14} we introduced a method to extract signatures of nuclear
deformation from auxiliary-field quantum Monte Carlo (AFMC) simulations in
the framework of the spherical configuration-interaction (CI) shell model.
The method works by examining the finite-temperature distributions of the
axial quadrupole operator $\hat{Q}_{20}$ in the laboratory frame. Here we
expand upon the details of the method, describing the calculation of this
distribution and how to overcome an equilibration and decorrelation problem
for deformed nuclei.  Our current methodology was demonstrated 
in Ref.~\cite{al14} for the spherical nucleus
$^{148}$Sm and the deformed nucleus $^{154}$Sm. Here we extend the
applications to isotope chains of even-mass samarium and neodymium nuclei
which exhibit a crossover from spherical to deformed shapes.

We use quadrupole invariants~\cite{ku72,cl86}, defined
in the framework of the CI shell model, to extract information on the
effective intrinsic quadrupole deformation.  This aspect is independent
of the method used to determine the invariants.  For nuclei lighter
than the ones considered here, the CI shell model has also been successfully
employed.  See Refs.~\cite{ha16,sc17} for recent examples and for other references cited therein.  We note the AFMC has been previously applied
to  calculate intrinsic shape distributions~\cite{al96}  but there an ad hoc prescription was used for extracting
shape information.

The outline of this paper is as follows. In Sec.~\ref{AFMC} we review briefly the finite-temperature AFMC method. In Sec.~\ref{q-project}, we discuss the formalism for projecting onto the axial quadrupole operator in order to calculate its lab-frame finite-temperature distribution using AFMC. Furthermore, we present an angle-averaging method to help equilibrating this distribution and reduce the decorrelation length in its sampling. In Sec.~\ref{application} we apply the method to the deformed nucleus $^{162}$Dy and to two isotope chains of lanthanide nuclei in which we observe a crossover from spherical to deformed behavior. We compare our AFMC results with the finite-temperature HFB mean-field approximation~\cite{go81,ta81}. In Sec.~\ref{invariants}, we discuss low-order quadruple invariants and their relation to the moments of the axial quadrupole operator in the laboratory frame. These invariants are used to extract effective intrinsic quadrupole deformation parameters in the rotationally invariant framework of the CI shell model, without the use of a mean-field approximation. We conclude with a summary and outlook in Sec.~\ref{summary}. Some of the technical details are discussed in the Appendices. Finally, the data files containing the AFMC and HFB results presented in this work are included in the Supplementary Material depository of this article.

\section{Auxiliary-field Monte Carlo method}\label{AFMC}

We briefly review the AFMC method, also known as shell model Monte Carlo (SMMC) in the context of the nuclear shell model~\cite{la93,al94,al17}. For a nucleus at a finite temperature $T$, we consider its imaginary-time propagator $e^{-\beta \hat H}$, which describes the Gibbs ensemble at inverse temperature $\beta = 1/T$ for a
Hamiltonian $\hat H$.\footnote{We will use the circumflex
to denote operators in the many-particle Fock space.}
The AFMC method is based on the Hubbard-Stratonovich
transformation~\cite{HS}, in which the Gibbs propagator $e^{-\beta \hat H}$ is decomposed into a
superposition of imaginary-time evolution operators $\hat U_\sigma$  of non-interacting nucleons
\be
  e^{-\beta \hat H} = \int \mathcal D [\sigma] \; G_\sigma  \hat U_\sigma \;,
\ee
where $G_\sigma$ is a Gaussian weight and $\sigma$ are auxiliary fields that depend on imaginary time $\tau$ ($0 \le \tau\le \beta$).

The thermal expectation value of an observable $\hat O$ is then given by
\begin{eqnarray}\label{observ}
  \langle \hat O \rangle=
          {{\rm Tr}\,( \hat O e^{-\beta \hat H})\over{\rm Tr}\,(e^{-\beta \hat H})} =
          {\int {\cal D}[\sigma] G_\sigma \langle \hat O \rangle_\sigma{\rm Tr}\,\hat U_\sigma
\over \int {\cal D}[\sigma] G_\sigma {\rm Tr}\,\hat U_\sigma}  \;,
\end{eqnarray}
 where  $\langle \hat O \rangle_\sigma\equiv
 {\rm Tr} \,( \hat O \hat U_\sigma)/ {\rm Tr}\,\hat U_\sigma$ is the expectation value of $\hat O$ for non-interacting particles in external auxiliary fields $\sigma(\tau)$.
 
The grand-canonical traces can be evaluated by reducing them to quantities in the single-particle space.  For example, the grand-canonical partition function for a given configuration of the fields $\sigma$ is 
 \be\label{partition}
{\rm Tr}\; \hat U_\sigma = \det ( {\bf 1} + {\bf U}_\sigma) \;,
\ee
where ${\bf U}_\sigma$ is the matrix representation of $\hat U_\sigma$ in the single-particle space.

The Monte Carlo sampling of the fields $\sigma$ is carried out using the positive-definite weight function
\be\label{W-function}
 W_\sigma \equiv
  G_\sigma \vert \Tr \;  \hat U_\sigma \vert \;.
\ee
We define the $W$-weighted average of a quantity $X_\sigma$ that depends on the auxiliary-field configuration $\sigma$ by 
\be\label{ave_x}
\left \langle X_\sigma \right\rangle_W \equiv \frac {\int D[\sigma ] W_\sigma  X_{\sigma} \Phi_{\sigma}} { \int D[\sigma]  W_\sigma \Phi_{\sigma}} \;,
\ee
where
\be \label{sign}
\Phi_\sigma\equiv \Tr\; U_\sigma /\vert \Tr \; U_\sigma \vert
\ee
 is the Monte Carlo sign function.  With this definition, the thermal expectation of an observable $\hat O$ can be written as 
  \be\label{observ-W}
 \langle \hat O \rangle =  \left\langle \Tr ( \hat O \hat U_\sigma) / \Tr  \hat U_\sigma \right\rangle_W  \;.
 \ee

We can also carry out projections on conserved one-body observables, such as particle number~\cite{or94,al99} and total angular momentum~\cite{al07}. In particular, the canonical partition $\Tr_{\cal A} \hat U_\sigma$ for fixed particle number ${\cal A}$ can be calculated by the discrete Fourier transform
\begin{eqnarray}\label{canonical}
{\rm Tr}_{\cal A} U_\sigma =\frac{e^{-\beta\mu {\cal A}}
}{N_s}\sum_{m=1}^{N_s}
e^{-i\chi_m {\cal A}}\det \left( {\bf 1}+e^{i\chi_m}e^{\beta\mu}{\bf U}_\sigma\right),
\end{eqnarray}
where $N_s$ is the number of single-particle orbitals, $\chi_m=2\pi m/N_s \;\; (m=1,\ldots,N_s)$ are quadrature points and $\mu$
is a chemical potential introduced to stabilize the numerical evaluation of the Fourier sum.

In practice, we always work in the canonical ensemble and calculate the expectation values of observables at fixed proton and neutron numbers, i.e., ${\cal A} = (Z,N)$. We then replace the traces in Eqs.~(\ref{observ},\ref{W-function}-\ref{observ-W}) by canonical traces, i.e., $\Tr \to \Tr_{\cal A}$. 

\section{Projection on the axial quadrupole operator $\hat Q_{20}$ in the laboratory frame}\label{q-project}

\subsection{Projection formalism}

The probability distribution of the axial quadrupole operator 
$\hat Q_{20} = \sum_i \left(2  z_i^2 -  x_i^2 - y_i^2 \right)$ at inverse temperature $\beta$ is defined by
\be\label{Pq}
P_\beta(q)  ={ {\Tr [\delta({\hat Q_{20}} - q) e^{-\beta \hat H}]}/\Tr e^{-\beta \hat H}} \;,
\ee
where $q \equiv q_{20}$ and $\delta({\hat Q_{20}} - q)$ projects onto the eigenspace corresponding to eigenvalue $q$ of $\hat{Q}$. The Monte Carlo expectation of $P_\beta(q)$ is then
\be\label{prob-HS}
P_\beta(q) = \left\langle \frac
{\Tr\left[\delta(\hat Q_{20} - q) \hat U_\sigma \right]}
{\Tr {\hat U}_\sigma} \right\rangle_W  \;.
\ee

Expanding the distribution (\ref{Pq}) in terms of many the many-particle eigenstates $|q_n \rangle$ and $|e_m\rangle$  of $\hat Q_{20}$ and $\hat H$, respectively, we find
\be\label{prob1}
P_\beta(q) = \sum_n \delta(q - q_n) \sum_m \langle q_n |e_m \rangle^2 e^{-\beta
e_m} \;.
\ee
$P_\beta(q)$ represents the probability of measuring eigenvalue $q$ of $\hat{Q}_{20}$ in the finite-temperature Gibbs ensemble. Since the quadrupole operator $\hat Q_{20}$ does not commute with the Hamiltonian, there does not exist a basis of simultaneous eigenstates of $\hat Q_{20}$ and $\hat H$, unlike for projections onto conserved one-body observables.  Nevertheless, the quadrupole distribution in a thermal ensemble is well-defined and is given by Eq.~(\ref{prob1}).

The expression (\ref{prob1}) is impractical for realistic calculations, since the sums over $n,m$ range over bases of many-particle states. However, the distribution $P_\beta(q) $ can be calculated using the Fourier representation of the $\delta$ function
\be \label{delta-q}
\delta(\hat Q_{20} - q) = {1 \over 2 \pi} \int_{-\infty}^\infty d \varphi \, e^{-i \varphi q }\, e^{i \varphi \hat Q_{20}} \,.
\ee

Up to now, we have treated $q$ as a continuous variable. Since the AFMC works in a finite model space, $\hat Q_{20}$ has a discrete and finite spectrum. However, for heavy nuclei, the many-body eigenvalues of $\hat Q_{20}$ are sufficiently closely spaced to allow $q$ to be approximated as continuous.

For a given $\hat U_\sigma$, the projection is carried out using
a discretized version of the Fourier decomposition in Eq.~(\ref{delta-q}). We
take an interval $[-q_{\rm max},q_{\rm max}]$ and divide it into $2M+1$ 
equal intervals of length $\Delta q=2q_{\rm
max}/(2M+1)$. We define $q_m=m \Delta q$, where
$m=-M,\ldots,M$, and approximate the quadrupole-projected trace in
(\ref{prob-HS}) by
\be\label{fourier-q}
\Tr\left[\delta(\hat Q_{20} - q_m) \hat U_\sigma \right] \!\!  \approx \!\! {1\over 2 q_{\rm max}}\! \! \sum_{k=-M}^M
  \!\!\!  e^{-i \varphi_k q_m} \Tr(e^{i \varphi_k \hat Q_{20}} \hat U_\sigma) \;,
\ee
where $\varphi_k = \pi k/q_{\rm max}$ ($k=-M,\ldots, M$). 

Since $\hat Q_{20}$ is a one-body operator and $\hat U_\sigma$ is a one-body propagator,
the grand-canonical many-particle trace on the r.h.s.~of Eq.~(\ref{fourier-q})
reduces to a determinant in the single-particle space 
\be
 \Tr\left(e^{i\varphi_k \hat Q_{20}} \hat U_\sigma \right) = 
  \det  \left( 1+   e^{i \varphi_k {\bf Q}_{20}} {\bf U}_\sigma\right) \;,
  \ee
where ${\bf Q}_{20}$ and ${\bf U}_\sigma$ are  the matrices representing, respectively, $\hat Q_{20}$ and $\hat U_\sigma$, in the single-particle space.  Projections are also carried on the  proton and proton number operators to fix the $Z$ and $N$ of the ensemble. 

\subsection{Angle averaging and thermalization}

The thermalization and decorrelation of moments of $\hat Q_{20}$ are very slow for deformed nuclei when using the pure Metropolis sampling.  This can be overcome by augmenting the Metropolis-generated configurations by rotating them through a properly chosen set of $N_{\Omega}$ rotation angles  $\Omega_i$.  In practice, it is easier to rotate the matrix ${\bf U}_\sigma$.  We make the replacement
\be 
\langle e^{i \varphi \hat Q_{20}} \rangle_\sigma \rightarrow {1\over
N_\Omega} \sum_{j=1}^{N_\Omega} \langle e^{i \varphi \hat Q_{20}} \rangle_{\sigma,\Omega_j} \;,
\ee
where
\be
\langle \hat X \rangle_{\sigma,\Omega} = \frac{\Tr \left [ \hat X \, \left (\hat R(\Omega) \hat U_\sigma \hat R^\dagger(\Omega) \right) \right ]}{\Tr \left [\hat R(\Omega) \hat U_\sigma \hat R^\dagger(\Omega) \right ]}\;,
\ee
and where $\hat R(\Omega)$ is the rotation operator for angle $\Omega$.

In the following we discuss methods for choosing these angles. The main observation is that quadrupole invariants such as $Q \cdot Q$  thermalize faster than moments of $\hat Q_{20}$. For a rotationally invariant system, the low moments of $\hat Q_{20}$ are proportional to the expectation values of quadrupole invariants (see Eqs.~(\ref{q202}-\ref{q204}) below).  In AFMC, these relations hold only on average, i.e., only after averaging over all auxiliary-field configurations. If we can choose a set of rotation angles $\Omega_i$ such the angle-averaged moment of $\hat Q_{20}$ is equal to the corresponding invariant (as operators), then the relations~(\ref{q202}-\ref{q204}) will hold sample-by-sample. We will show that this leads to a faster thermalization and decorrelation of the corresponding moment of $\hat Q_{20}$.

\subsubsection{Six-angle averaging} \label{sixangles}

Here, we find a set of six angles for which the angle-averaged $\hat Q_{20}^2$ is proportional to the second-order invariant $\hat Q \cdot \hat Q$.

The second moment of $\hat Q_{20}$ is related to the second-order quadrupole invariant $\langle \hat Q \cdot \hat Q \rangle$ by $\langle \hat Q^2_{20} \rangle = \langle \hat Q \cdot \hat Q \rangle/5$. We would like to find a set of angles $\Omega_i$ such that
\be
\sum_i \hat{R}(\Omega_i) \hat{Q}_{2 0}^2 \hat{R}^{\dagger}(\Omega_i) \propto \hat{Q} \cdot \hat{Q} \,.
\ee

\begin{figure*}[t!]
 \includegraphics[angle=0,width=0.7\textwidth]{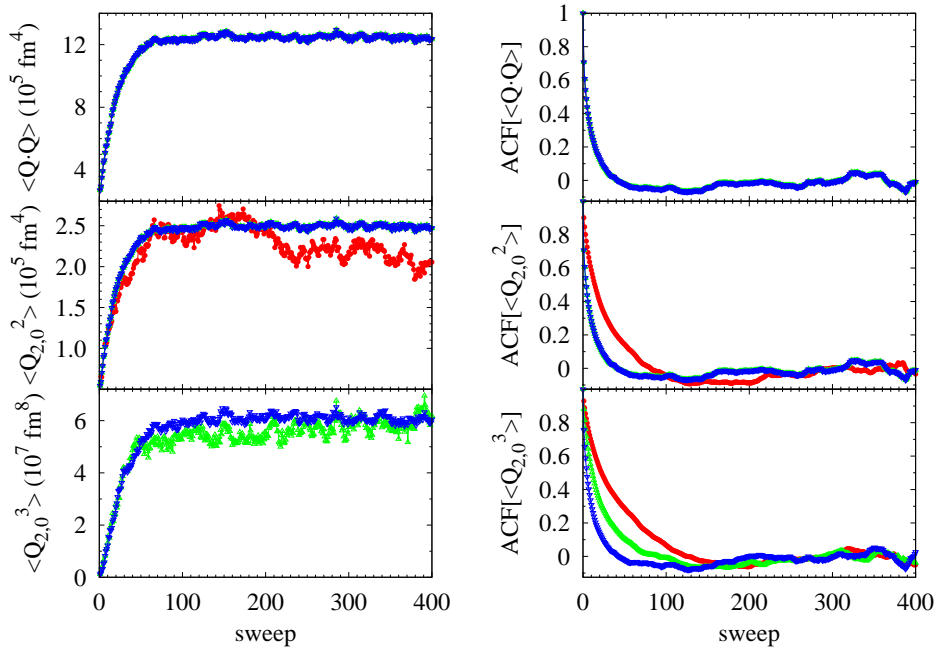}
 \caption{Equilibration (left column) and autocorrelation function (right column) of the  invariant $\langle \hat{Q} \cdot \hat{Q} \rangle$ (top panels), and of the quadrupole moments $\langle \hat{Q}_{20}^2 \rangle$ and $\langle \hat{Q}_{20}^3 \rangle$, for several levels of angle-averaging: no averaging (red circles), six angles (green triangles), and 21 angles (blue inverted triangles). The horizontal axis shows the number of Monte Carlo sweeps (each sweep is one update of all auxiliary fields). In the top panels, the points overlap since $\langle \hat{Q} \cdot \hat Q\rangle$ is rotationally invariant sample-by-sample. In the bottom right panel, the no-averaging results have $\langle \hat{Q}_{20}^3\rangle \sim 10^{10}$ and are above the range of the y-axis. Rotating by a sufficient number of angles to restore rotational invariance for $\hat Q_{20}^2$ and $\hat Q_{20}^3$ makes their Monte Carlo sampling behavior become very similar to that of $\langle \hat Q \cdot \hat Q \rangle$.}
 \label{sm_qtherm}
\end{figure*} 

To determine an appropriate set of angles, we rotate each factor of $\hat Q_{20}$ to obtain
\begin{multline}
\sum_i  \hat{R}(\Omega_i) \hat{Q}_{2 0}^2 \hat{R}^{\dagger}(\Omega_i) \\ =  \sum_{\mu,\mu'}  \left(  \sum_i D^{(2)}_{\mu 0}(\Omega_i) D^{(2)}_{\mu' 0}(\Omega_i)  \right)  \hat{Q}_{2 \mu} \hat{Q}_{2 \mu'} \;,
\end{multline}
where $D^{(2)}(\Omega)$ is the corresponding Wigner matrix for a rotation angle $\Omega$.  
The required condition is thus
\be\label{delta}
\sum_i D^{(2)}_{\mu 0}(\Omega_i) D^{(2)}_{\mu' 0}(\Omega_i) \propto (-)^{\mu} \delta_{\mu,-\mu'} \,.
\ee
To find the appropriate angles for which~\eqref{delta} holds, note that $D^{(2)}_{\mu 0}(\varphi,\theta,\psi) = \sqrt{\frac{4\pi}{5}} Y_{2 \mu}(\theta,\varphi) \propto P_{2}^{0}(\cos \theta) e^{i \mu \varphi}$. The Kronecker delta $\delta_{\mu,-\mu'}$ can be obtained by summing over five angles $\varphi_k = \frac{2\pi k}{5}$, $k=0,1,\ldots,4$ using the identity $\sum_{k=0}^{4} e^{2 \pi i k (\mu+\mu')/ 5} = 5 \delta_{\mu,-\mu'}$. Then
\be
\sum_{k=0}^4 D^{(2)}_{\mu 0}(\theta,\varphi_k) D^{(2)}_{\mu' 0}(\theta,\varphi_k) = 5 \delta_{\mu,-\mu'} P_2^\mu(\cos \theta) P_2^{-\mu}(\cos \theta)
\ee
for any $\theta$.  Choosing a particular angle $\theta_1$ by $\cos^2 \theta_1 = 1/5$, we have
\be
P_2^\mu(\cos \theta_1) P_2^{-\mu}(\cos \theta_1) = \begin{cases}
  1/5, & \mu = 0 \\
  -6/5, & \mu = \pm 1\\
  6/5, & \mu = \pm 2
\end{cases} \,.
\ee
Adding the angle $\theta=0, \varphi=0$ at which $D_{\mu 0}^{(2)}(0,0) = \delta_{\mu,0}$, we obtain
\be
\sum_{i=0}^{5} D^{(2)}_{\mu 0}(\Omega_i) D^{(2)}_{\mu' 0}(\Omega_i) = \frac{6}{5} (-)^\mu \delta_{\mu,-\mu'} \;,
\ee
where $\Omega_0 = (0,0)$ and $\Omega_i = (\theta_1, \varphi_{i-1})$ for $i=1,\ldots,5$.

In Appendix A, we discuss a general method to determine a set of angles such that the angle-averaged moments of $\hat Q_{20}$ are proportional to the quadrupole invariants of the same order up to the $n$-th order.  For $n=2$ and $n=3$, this leads to a set of 10 and 21 angles, respectively. When averaging over the specific set of 21 angles, both the second and cubic moments of $\hat Q_{20}$ are proportional to the second- and third-order quadrupole invariants, respectively, sample-by-sample.  

We demonstrate the thermalization of moments of $\hat Q_{20}$ for $^{154}$Sm in the left column Fig.~\ref{sm_qtherm}, in which we show the second moment $\langle \hat Q^2_{20}\rangle$ (middle panel) and third moment  $\langle \hat Q^3_{20}\rangle$ (bottom panel) of $\hat Q_{20}$  as a function of the sweep number in the Monte Carlo random walk.  We compare the no angle averaging (red circles) with the 6-angle averaging (green triangles) and 21-angle averaging (blue inverted triangles). The top panel shows the direct calculation of the second-order quadrupole invariant $\langle \hat{Q} \cdot \hat{Q} \rangle$, which has no equilibration or decorrelation problem. One can see that $\langle \hat Q_{20}^2 \rangle$ for 6 and 21 angles, and $\langle \hat Q_{20}^3 \rangle$ for 21 angles, show smaller fluctuations and more rapid and obvious thermalization, very similar to the behavior of $\langle \hat Q \cdot \hat Q \rangle$.

On the right column of Fig.~\ref{sm_qtherm}, we show the autocorrelation function (ACF) of the same observables and for similar levels of angle-averaging.  The observable $\langle \hat{Q}_{20}^2 \rangle$ decorrelates faster in the six-angle averaging (as compared with the no-angle averaging results), while $\langle \hat{Q}_{20}^3 \rangle$ decorrelates fastest in the 21-angle averaging. We see that the improvement in thermalization is closely related to a shorter decorrelation length.

\begin{figure}[bth]
   \includegraphics[angle=0,width=0.8\columnwidth]{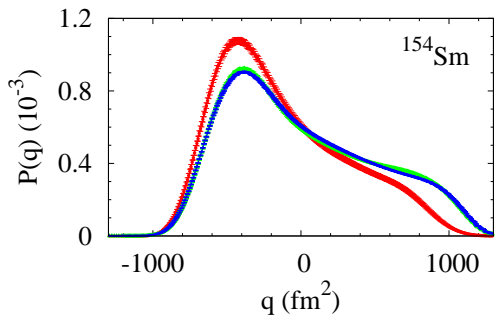}
    \caption{Distribution $P(q)$ vs.~$q$ of $^{154}\text{Sm}$ at a low temperature for several levels of angle-averaging: no averaging (red symbols), six angles (green symbols), and 21 angles (blue symbols). The results from 6-angle and 21-angle averaging are hardly distinguishable. }
    \label{sm_zvq_avging}
\end{figure} 

The thermalization of the second and third moments of $\hat Q_{20}$ is important for the calculation of a distribution $P(q)$ that is independent of the choice of the initial seed. In Fig.~\ref{sm_zvq_avging} we show the quadrupole distribution for $^{154}$Sm using different levels of angle-averaging, based on $\sim 5000$ Monte Carlo samples.  In the  AFMC calculations, it is necessary to discretize the imaginary time and we use a time slice of $\Delta \beta=1/32$ MeV$^{-1}$ here and in all the results presented in this work. The distribution obtained with no angle averaging (red line) is not thermailized, while the distribution with the six-angle average (blue line) is very close to the one obtained with 21-angle average (green line). In the applications presented in this work, we have used 21-angle averaging to make sure the calculated distributions $P(q)$ and moments of $\hat Q_{20}$ are thermalized and have a short decorrelation length. 

\section{Application to lanthanides}\label{application}

Here we present results for rare-earth nuclei. We consider the deformed nucleus $^{162}$Dy as well as the two isotope chains of even-even nuclei $^{144-152}$Nd and $^{148-154}$Sm.

 We compare some of the AFMC results with a mean-field approximation, the finite-temperature HFB approximation~\cite{go81,ta81}, using the same model space and interaction as for our AFMC calculations. In the zero-temperature HFB, there is a phase transition from a spherical to a deformed shape as we increase the number of neutrons within each of these two isotope chains. In the finite-temperature HFB, an isotope which has a deformed HFB ground state undergoes a phase transition to a spherical shape at a certain critical temperature. 
 
\subsection{CI shell model space and interaction}
 
The single-particle orbitals and energies are taken as eigenfunctions of a spherical Woods-Saxon potential plus a
spin-orbit interaction. For protons we take the complete 50-82 
shell plus the $1f_{7/2}$ orbital, and for neutrons the $82-126$ shell plus
the $0h_{11/2}$ and $1g_{9/2}$ orbitals.  The interaction includes a monopole pairing interaction and multipole-multipole interactions with quadrupole, octupole and hexadecupole components.  The interaction parameters are given in Refs.~\cite{al08,oz13}.

\subsection{Distributions $P_\beta(q)$}

\subsubsection{ A deformed nucleus}

\begin{figure}[htb]
    \includegraphics[angle=0,width= 0.65\columnwidth]{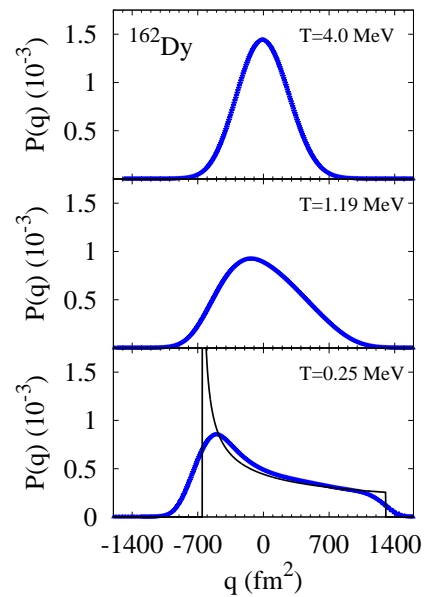}
    \caption{Distributions $P(q)$ vs.~$q$ for $^{162}$Dy at high temperature (top panel), near the HFB shape transition temperature (middle panel), and at low temperature (bottom panel). The low-temperature distribution is qualitatively similar to the rigid rotor ground-state distribution (solid line) with an intrinsic quadrupole moment $q_0$ taken to be the HFB value of $\langle \hat{Q}_{2 0}\rangle$. }
\label{dy162_zvq_temp}
\end{figure} 

In Fig.~\ref{dy162_zvq_temp} we show the distributions $P_\beta(q)$ for $^{162}$Dy at high temperature, near the HFB shape transition temperature, and at low temperature.  Since our model space is restricted to valence shells, we scale $q$  here and in other figures by a factor of 2 to account for effects of the core. One expects the nucleus to resemble a rigid rotor at low temperatures. For a rigid rotor with an intrinsic quadrupole moment of $q_0>0$ (corresponding to a prolate shape), this distribution is~\cite{al14}
\begin{eqnarray}\label{prob-rotor}
P_{\rm g.s.}(q) =  \left\{ \!\!\! \begin{array}{cl}  
\left(\sqrt{3} q_0 \sqrt{1 +2 {q\over q_0}}\right)^{-1}  & \! \! \mbox{for $-{q_0 
\over 2} \leq q \leq q_0$}  \\
 0  & \mathrm{otherwise}  \end{array}  \right. \;.
 \end{eqnarray}
We determine $q_0$ from the HFB value of $\langle \hat{Q}_{20}\rangle$ at $T=0$. The rigid-rotor distribution shown in the bottom panel of Fig.~\ref{dy162_zvq_temp} is qualitatively similar to the low-temperature distribution for $^{162}$Dy.

The lab-frame moments of the distribution (\ref{prob-rotor}) are calculated in Appendix B. 

\subsubsection{A spherical nucleus} 

\begin{figure}[h!]
    \includegraphics[angle=0,width=0.65\columnwidth]{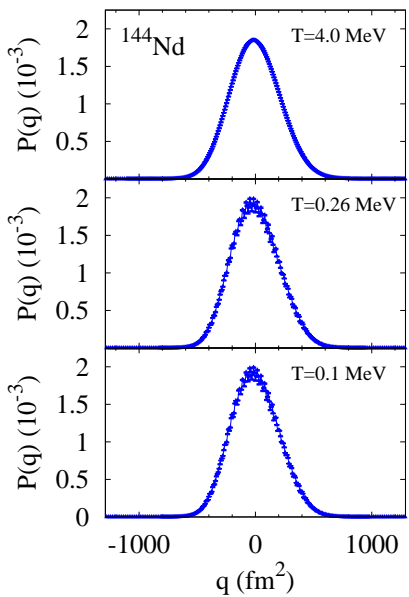}
    \caption{Distributions $P(q)$  vs $q$ for $^{144}$Nd at high temperature (top panel), an intermediate temperature (middle panel), and at low temperature (bottom panel). The low-temperature distributions show a slight staggering effect and skew but are very close to Gaussian.}
    \label{nd144_zvq_temp}
\end{figure}

In Fig.~\ref{nd144_zvq_temp} we illustrate the case of a spherical nucleus, $^{144}$Nd. Here the quadrupole distribution is close to Gaussian at all temperatures, although it develops a slight skew at low temperatures.
 
We conclude that the distribution of the axial quadrupole in the laboratory frame is a model-independent signature of deformation.

\begin{figure*}[htb]
  \includegraphics[angle=0,width=\textwidth]{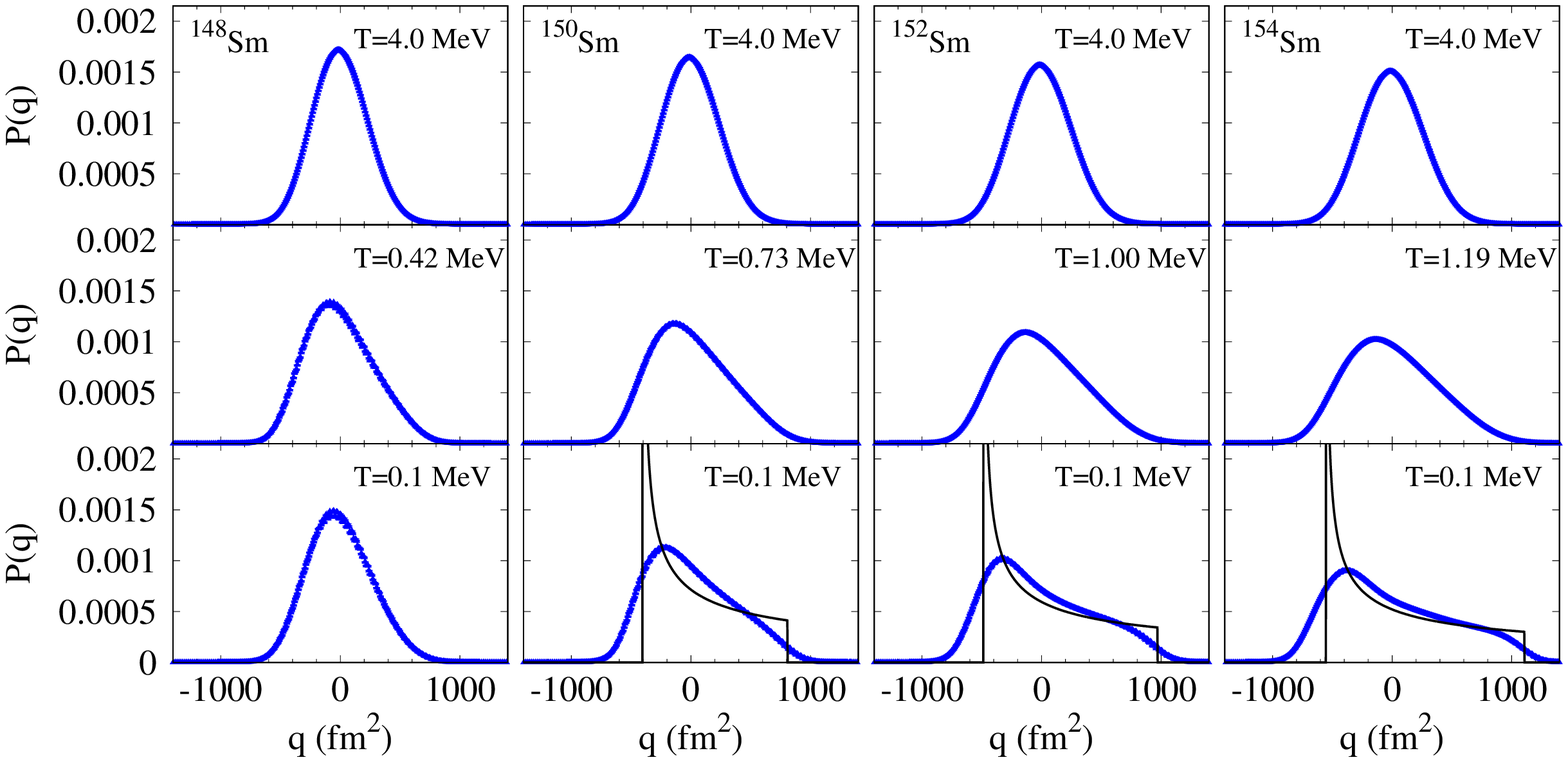}
    \caption{Distributions $P(q)$ vs.~$q$ for an isotope chain of even-mass samarium nuclei at high temperature (top row), intermediate temperatures (middle row), and at low temperature (bottom row). In the middle row, the deformed nuclei ($^{150}\text{Sm}$, $^{152}\text{Sm}$, $^{154}\text{Sm}$) are shown near their HFB shape transition temperatures. The solid lines are the rigid-rotor distributions (\ref{prob-rotor}) with the ground-state HFB values of  the intrinsic quadrupole moment $q_0$.}
    \label{sm_zvq_isotopes}
\end{figure*} 
\begin{figure*}[htb]
  \includegraphics[angle=0,width= \textwidth]{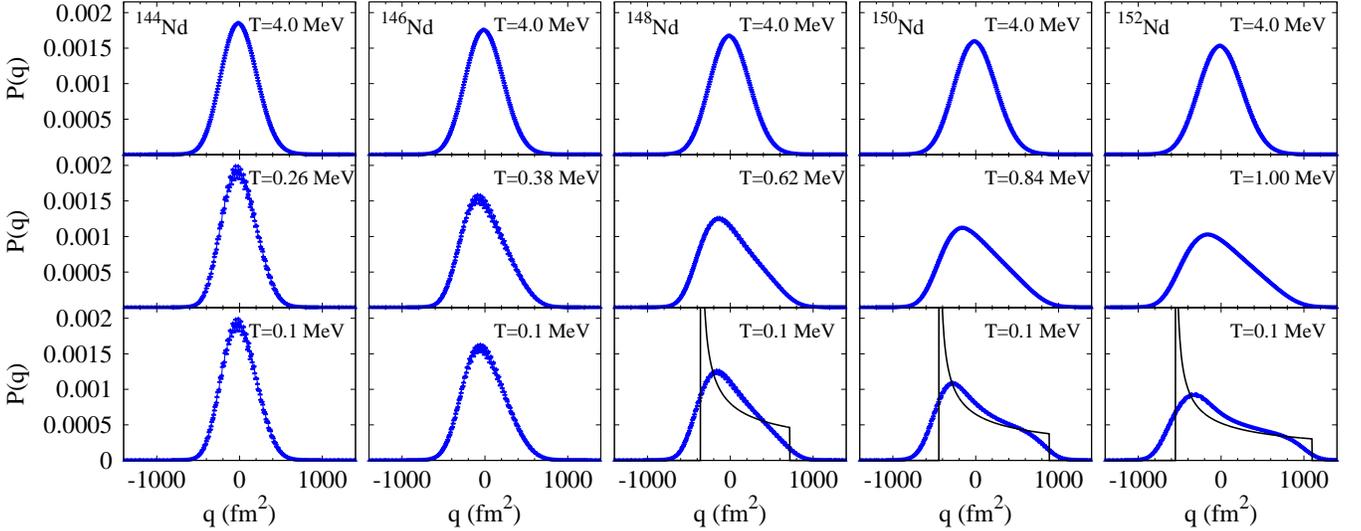}
    \caption{As in Fig.~\ref{sm_zvq_isotopes} but for an isotope chain of even-mass neodymium nuclei. In the middle row, the deformed nuclei ($^{148}\text{Nd}$, $^{150}\text{Nd}$, $^{152}\text{Nd}$) are shown near their HFB shape transition temperatures.}
    \label{nd_zvq_isotopes}
\end{figure*}

\subsubsection{Crossover from spherical to deformed nuclei}

In Figs.~\ref{sm_zvq_isotopes} and ~\ref{nd_zvq_isotopes} we show the distributions $P_\beta(q)$ for isotope chains of samarium and neodymium nuclei at several temperatures. The low-temperature ($T=0.1$ MeV) distributions display a crossover as a function of neutron number from spherical behavior in $^{148}$Sm and $^{144}$Nd to deformed behavior in $^{154}$Sm and $^{152}$Nd.

The nuclei that are deformed at low temperatures undergo a crossover to a spherical shape with increasing temperature. This transition from deformed to spherical shape is well-known in the HF and HFB mean-field theories where it is seen as a sharp phase transition~\cite{go86,ma03,ag00}. Here we see it as a gradual change. The distributions $P_\beta(q)$ are still skewed above the mean-field transition temperature, indicating the persistence of deformation effects to high temperatures. 

\subsection{ $\langle \hat Q \cdot \hat Q\rangle$ vs. temperature}

We now compare in more detail the AFMC results to those of the finite-temperature HFB approximation. The HFB solution is described by temperature-dependent one-body density matrix $\boldsymbol \rho_\beta$ and pairing tensor $\boldsymbol \kappa_\beta$.  The second-order invariant $\langle \hat Q\cdot \hat Q\rangle$ is derived in HFB 
using Wick's theorem, resulting in
\begin{eqnarray}\label{Q^2_HFB}
\langle \hat Q\cdot \hat Q\rangle = q_0^2 + \sum_\mu (-)^\mu \tr\left[{\bf Q}_{2\mu}\, ({\bf 1}-\boldsymbol \rho_\beta)\, {\bf Q}_{2 -\mu}\, \boldsymbol \rho_\beta\right] \nonumber \\ + \sum_\mu (-)^\mu \tr\left[{\bf Q}_{2\mu}\, \boldsymbol \kappa_\beta \,{\bf Q}^T_{2 -\mu}\, \boldsymbol \kappa_\beta^\ast\right] \;.
\end{eqnarray}
Here $ q_0 \equiv \tr ({\bf Q}_{20} \boldsymbol \rho_\beta)$  is the intrinsic axial quadrupole moment.  

\begin{figure*}[htb]
  \includegraphics[angle=0,width=0.9\textwidth]{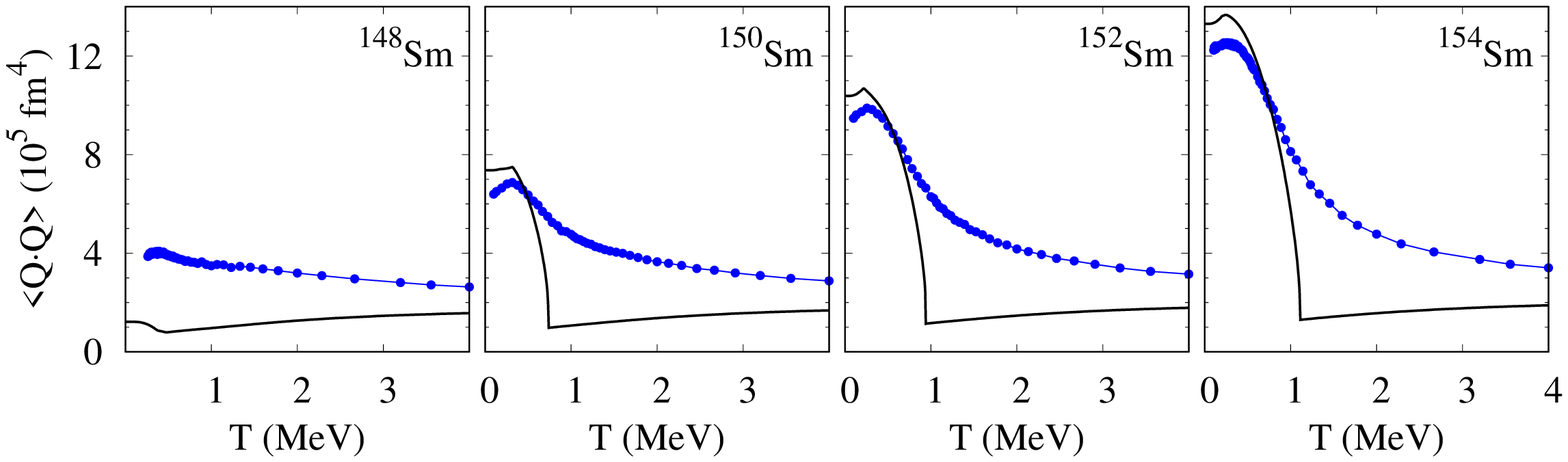}
    \caption{$\langle \hat Q \cdot \hat Q\rangle$ for Sm isotopes. The AFMC results (solid circles) are compared with the HFB results (solid lines).}
    \label{sm_qsqvt}
\end{figure*} 

\begin{figure*}[htb]
  \includegraphics[angle=0,width= \textwidth]{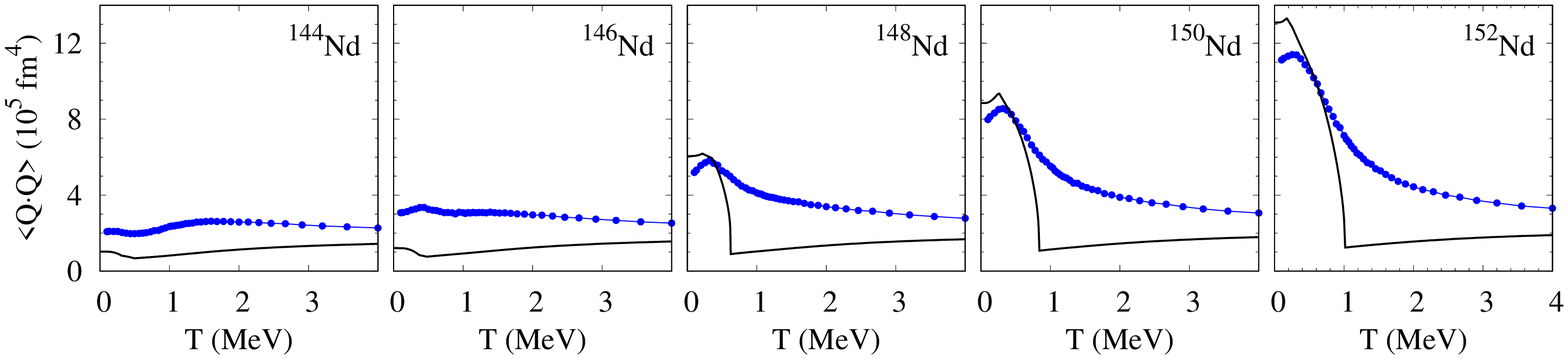}
    \caption{$\langle \hat Q \cdot \hat Q\rangle$ for Nd isotopes. Symbols and lines are as in Fig.~\ref{sm_qsqvt}.}
    \label{nd_qsqvt}
\end{figure*} 

In Figs.~\ref{sm_qsqvt} and~\ref{nd_qsqvt} we compare AFMC (blue solid circles) and HFB (solid lines) results for $\langle \hat Q \cdot \hat Q\rangle$ for the same isotope chains of samarium and neodymium nuclei. The HFB results show a sharp phase transition between spherical and deformed shapes as a function of temperature, while the AFMC shows a smooth crossover, as remarked earlier. For the deformed nuclei, $\langle \hat Q \cdot \hat Q\rangle$ is similar for the two methods at low temperature. 

\section{Quadrupole invariants and moments of $\hat Q_{20}$}\label{invariants}

While the axial quadrupole distribution in the laboratory frame exhibits a model-independent signature of deformation, the physical quantity of interest is the intrinsic deformation.  The intrinsic quadrupole deformation parameters are usually extracted in the framework of a mean-field approximation, and the challenge is to extract them in the rotationally invariant framework of the CI shell model. Combinations of the quadrupole operators $\hat Q_{2\mu}$ that are invariant under rotations, known as quadrupole invariants~\cite{ku72,cl86}, have the same values in both the laboratory frame and the intrinsic frame, and thus can provide information on the effective values of the intrinsic deformation parameters without resorting to a mean-field approximation. 

We define the low-order quadrupole invariants below and show that they are related to moments of $\hat Q_{20}$ of the same order in the following subsection \ref{moments}. 

\subsection{Quadrupole invariants}\label{quad_invariants} 

The quadrupole invariants can be classified by their order. The lowest-order invariant is quadratic
\be
\hat Q\cdot \hat Q = \sum_\mu (-)^\mu \hat Q_{2\mu} \hat Q_{2 -\mu} \;.
\ee
The third-order invariant is given by
\be
(\hat Q \times \hat Q ) \cdot \hat Q \! = \! \sqrt{5} \!\! \sum_{\mu_1,\mu_2,\mu_3} \!\! \left( \! \begin{array}{ccc}
 2 & 2 & 2 \\ \mu_1 &  \mu_2 & \mu_3  \end{array} \!\right) \! \hat Q_{2\mu_1}\hat Q_{2\mu_2}
\hat Q_{2\mu_3} \;.
\ee

There are three different ways to construct a fourth-order invariant. Under certain conditions, the fourth-order invariants are all proportional to $(\hat Q \cdot \hat Q)^2$
\be
\label{q4}
 (\hat Q \times \hat Q)^{(J)} \cdot (\hat Q \times \hat Q)^{(J)}  =
\begin{cases}
  \frac{1}{5}  (\hat Q \cdot \hat Q)^2 , & J = 0 \\
  \frac{2}{7}  (\hat Q \cdot \hat Q)^2 , & J = 2 \\
  \frac{18}{35}  (\hat Q \cdot \hat Q)^2 , & J = 4
\end{cases} \,,
\ee
which will be derived below. Similarly, the fifth-order invariant is also unique and we define it to be $(\hat Q \cdot \hat Q) ((\hat Q \times \hat Q ) \cdot \hat Q)$.

To derive Eq.~\eqref{q4},  we will assume that the quadrupole operators $\hat Q_{2\mu}$ commute among themselves. This holds for the quadrupole operators in coordinate space but not in the truncated CI shell model space. We believe the effect of their noncommutation is small and so we will ignore this in the following. Working in basis of simultaneous eigenstates of $\hat{Q}_{2\mu}$ with eigenvalues $q_{2 \mu}$, we then rotate to an intrinsic frame, in which we will denote the quadrupole components by ${\tilde q}_{2 \mu}$. This frame is defined by the conditions
\be
{\tilde q}_{21} = {\tilde q}_{2\,-1} = 0, \;\;\; {\tilde q}_{22} = {\tilde q}_{2\,-2} = \text{real} \,.
\ee
To calculate the fourth-order invariants, we expand
\be
({\tilde q} \times {\tilde q})^{(J)} \cdot ({\tilde q} \times {\tilde q})^{(J)} = \sum_\mu (-)^\mu ({\tilde q} \times {\tilde q})^{(J)}_\mu ({\tilde q} \times {\tilde q})^{(J)}_{-\mu}\,
\ee
as well as $({\tilde q} \times {\tilde q})^{(J)}_\mu = \sum \cg{2}{m}{2}{\mu-m}{J}{\mu} {\tilde q}_{2 m} {\tilde q}_{2 \mu-m}$. Only terms with even $\mu$ contribute to the sum in this frame. Evaluating the Clebsch-Gordan coefficients and simplifying, one obtains
\be
({\tilde q} \times {\tilde q})^{(J)} \cdot ({\tilde q} \times {\tilde q})^{(J)} =
\begin{cases}
  \frac{1}{5} ({\tilde q}_{20}^2 + 2 {\tilde q}_{22}^2)^2, & J = 0 \\
  \frac{2}{7} ({\tilde q}_{20}^2 + 2 {\tilde q}_{22}^2)^2, & J = 2 \\
  \frac{18}{35} ({\tilde q}_{20}^2 + 2 {\tilde q}_{22}^2)^2, & J = 4 .
\end{cases}
\ee
Since $({\tilde q}_{20}^2 + 2 {\tilde q}_{22}^2) = {\tilde q} \cdot {\tilde q} = q \cdot q$ in this frame, we obtain Eq.~\eqref{q4} by rotational invariance.

\subsection{Relations of the quadrupole invariants to moments of $\hat Q_{20}$}\label{moments}

When the invariant  is unique at a given order, its expectation value can be computed directly from the corresponding lab-frame moment of $\hat Q_{20}$, defined by  $\langle \hat Q_{20}^n \rangle_\beta =  \int  q^n P_\beta(q)d q$. For a rotationally invariant system, the expectations $\langle \hat{Q}^n_{20}\rangle$ for $n=2,3,4$ are related to the invariants by~\cite{al14}
\begin{eqnarray}
   \langle \hat{Q} \cdot \hat{Q} \rangle & = & 5 \langle \hat{Q}_{20}^2 \rangle \;, \label{q202}\\
 \langle (\hat{Q} \times \hat{Q})^{(2)} \cdot \hat{Q} \rangle & = &   - 5 \sqrt{\frac{7}{2}} \langle \hat{Q}_{20}^3 \rangle \;, \label{q203}\\
  \langle (\hat{Q} \cdot \hat{Q})^2 \rangle & = &  \frac{35}{3} \langle \hat{Q}_{20}^4 \rangle \;,\label{q204}\\
 \langle (\hat Q \cdot \hat Q) ((\hat Q \times \hat Q ) \cdot \hat Q) \rangle & = &  -\frac{11}{2} \sqrt{\frac{7}{2}} \langle \hat{Q}_{20}^5 \rangle \;.\label{q205}
\end{eqnarray}
We now derive Eqs.~(\ref{q202}-\ref{q204}). For Eq.~\eqref{q202}, note that $\langle \hat{Q} \cdot \hat{Q} \rangle = \sum_\mu (-)^{\mu} \langle \hat{Q}_{2 \mu} \hat{Q}_{2  -\mu} \rangle = \sum_{\mu} \langle \hat{Q}_{2 \mu}^\dagger \hat{Q}_{2 \mu} \rangle$, since $\hat{Q}_{2 \mu}^\dagger = (-)^\mu \hat{Q}_{2 -\mu}$ (i.e., $\hat Q_{2\mu}$ is an hermitian operator). But for a rotationally invariant system, $\langle \hat{T}^{(J)}_{M} \hat{T}^{(J')\dagger}_{M'}\rangle \propto \delta_{J,J'} \, \delta_{M,M'}$ and is independent of $M$ for any spherical tensor operator $\hat{T}^{(J)}$. This leads to relation (\ref{q202}).

For the third moment, write
\begin{align}
  \langle \hat{Q}_{20}^3 \rangle & = \sum_{J} \cg{2}{0}{2}{0}{J}{0} \langle (\hat{Q} \times \hat{Q})^{(J)}_0 \hat{Q}_{20}\rangle \\
  & = \sum_{J, K} \cg{2}{0}{2}{0}{J}{0}  \cg{J}{0}{2}{0}{K}{0} \langle(\hat{Q} \! \times \! \hat{Q})^{(J)} \! \times \! \hat{Q})^{(K)}_{0}\rangle .
\end{align}
Due to rotational invariance only the $K=0$ term contributes, which also fixes $J=2$. Using $\cg{2}{0}{2}{0}{2}{0} = -\sqrt{2/7}$ and $\cg{2}{0}{2}{0}{0}{0} = 1/\sqrt{5}$, we obtain Eq.~\eqref{q203}.

The fourth moment of $\hat Q_{20}$ can be calculated in a similar manner by writing
\begin{multline}
  \langle \hat{Q}_{20}^4 \rangle = \sum_{J,J',K} \cg{2}{0}{2}{0}{J}{0} \cg{2}{0}{2}{0}{J'}{0} \cg{J}{0}{J'}{0}{K}{0} \\
  \times \langle [(\hat{Q} \times \hat{Q})^{(J)} \times (\hat{Q} \times \hat{Q})^{(J')}]^{(K)}_{0}\rangle \,.
\end{multline}
Again only $K=0$ contributes to the sum, requiring $J=J'$. Also $\cg{2}{0}{2}{0}{J}{0} \neq 0$ only for $J=0,2,4$. Noting that $(\hat{T}^{(J)} \times \hat{T}^{(J)})^{(0)}_0 = (-)^{J} (\hat{T}^{(J)} \cdot \hat{T}^{(J)})/\sqrt{2J+1}$ and $\cg{J}{0}{J}{0}{0}{0} = (-)^{J}/\sqrt{2J+1}$, we obtain

\be
  \langle \hat{Q}_{20}^4 \rangle = \sum_{J} \frac{\cg{2}{0}{2}{0}{J}{0}^2}{2J+1} \langle (\hat{Q} \times \hat{Q})^{(J)} \cdot (\hat{Q} \times \hat{Q})^{(J)}\rangle \,.
\ee

Expressing the fourth-order invariants $\langle (\hat{Q} \times \hat{Q})^{(J)} \cdot (\hat{Q} \times \hat{Q})^{(J)}\rangle$ in terms of $(\hat Q \cdot \hat Q)^2$ using Eq.~\eqref{q4}, we find the result in Eq.~(\ref{q204}).  

Relation (\ref{q205}) for the fifth-order invariant can be derived in a similar manner.

\subsection{Effective deformation parameters}

We now use the quadrupole invariants to define effective deformation parameters which can be calculated from quantities known only in the lab frame.
We define quadrupole deformation parameters $\alpha_{2\mu} $ using a liquid drop model for which~\cite{ri80}
\be
q_{2\mu} = \frac{3}{\sqrt{5 \pi}} 3 r_0^2 A^{5/3} \alpha_{2\mu}\;.
\ee

The quadrupole deformation parameters in the intrinsic frame $\tilde\alpha_{2\mu}$ can be parametrized by the intrinsic parameters $\beta,\gamma$ of the collective Bohr Hamiltonian (see Sec.~6B-1a of Ref.~\cite{BM75})
\be
{\tilde\alpha}_{20}=\beta\cos \gamma \;;\;\;  {\tilde\alpha}_{22}={\tilde\alpha}_{2-2}=\frac{1}{\sqrt{2}}\beta\sin\gamma\;.
\ee
We can write the quadrupole invariants in terms of ${\tilde\alpha}_{20}$ and ${\tilde\alpha}_{22}$ and then express them in terms of $\beta, \gamma$.  The three lowest-order invariants are then given by $\beta^2, \beta^3\cos (3\gamma)$ and $\beta^4$. 

The second- and third-order invariants can be used to define effective values of the intrinsic
shape parameters $\beta,\gamma$
\be\label{effective_bc}
\beta = \frac{\sqrt{5 \pi}}{ 3 r_0^2 A^{5/3} } \langle \hat Q \cdot \hat Q \rangle^{1/2}  \;;\;\;
\cos 3\gamma = -\sqrt{7 \over 2} {\langle (\hat Q \times \hat Q ) \cdot \hat Q \rangle \over  \langle \hat Q \cdot \hat Q \rangle^{3/2} } \;.
\ee
We can also define an effective  fluctuation $\Delta \beta$ in $\beta$ from
\be\label{effective_db}
\left({\Delta \beta / \beta}\right)^2 =  {\left[\langle (\hat Q \cdot \hat Q)^2 \rangle - \langle \hat Q \cdot \hat Q \rangle^2\right]^{1/2} / \langle \hat Q \cdot \hat Q \rangle} \;.
\ee

\begin{table}[h]
  \begin{ruledtabular}\label{Table1}
    \begin{tabular}{c | c c c | c c c}
              & \multicolumn{3}{c|}{AFMC} & \multicolumn{3}{c}{HFB/5DCH} \\
      Nucleus & $\beta$ & $\Delta \beta / \beta$ & $\gamma$ (degrees)& $\beta$ & $\Delta \beta / \beta$ & $\gamma$ (degrees)\\
      \hline
      $^{144}$Nd & 0.106 & 0.755 & 25.0 & 0.118  &  0.29  &  28.  \\
      $^{146}$Nd & 0.126 & 0.744 & 22.0 & 0.167  &  0.26  &  25.
 \\
      $^{148}$Nd & 0.160 & 0.671 & 17.1  & 0.218  &  0.23  &  20.
\\
      $^{150}$Nd & 0.194 & 0.583 & 15.0  & 0.280  &  0.22  &  14.
\\
      $^{152}$Nd & 0.223 & 0.531 & 14.9  & 0.329  &  0.16  &  10.
\\
      $^{148}$Sm & 0.133 & 0.737 & 22.5  & 0.169  &  0.27  &  25.
 \\
      $^{150}$Sm & 0.173 & 0.627 & 16.2 & 0.229  &  0.25  &  20.
 \\
      $^{152}$Sm & 0.206 & 0.559 & 13.9 & 0.306  &  0.21  &  13.
 \\
      $^{154}$Sm & 0.230 & 0.520 & 13.7  & 0.342  &  0.15  &  10.
\\
    \end{tabular}
  \end{ruledtabular}
  \caption{Effective deformation parameters $\beta,\gamma$ and 
relative fluctuation $\Delta \beta/\beta$  for the even-mass nuclei 
$^{144-152}$Nd and $^{148-154}$Sm computed from the AFMC. 
Statistical errors are approximately $\pm 2$ in the last digit 
displayed. We used the 21-angle averaging AFMC results at $T=0.1$ MeV.
The last three columns on the right-hand-side show the corresponding quantities calculated
in an HFB model extended to include some fluctuations about the HFB
ground state~\cite{de10}. }
\end{table}

Table \ref{Table1} shows the effective values of $\beta$ and $\gamma$ calculated for the two isotope chains of even-mass samarium and neodymium nuclei using Eqs.~(\ref{effective_bc}) and (\ref{effective_db}).  Within each isotope chain, the effective values of $\beta$ increase with neutron number as the nucleus becomes more deformed. The nuclei within each isotope chain also become more rigid as indicated by the decrease of $\Delta \beta/\beta$. The respective values of the effective $\gamma$ decrease, being closer to triaxiality for the spherical nuclei and closer to axiality for the deformed nuclei.

The table also shows the corresponding quantities calculated in a
model based on self-consistent mean field theory but including
fluctuations in the deformation parameters~\cite{de10}. The trends
are all the same as in the AFMC results, and the triaxiality parameters agree fairly well
for these two very different theories.  The $\beta$ parameters are
systematically smaller in the AFMC. The table also
shows that the
fluctuation $\Delta \beta /\beta$ is significantly smaller in the HFB/5DCH than
in the AFMC.  This is also to be expected.  The degree of
freedoms that can fluctuate in the HFB/5DCH are very limited, 
while all the nucleonic degrees of freedom in the valence 
shells can participate in the AFMC fluctuations.

\section{Summary and outlook}\label{summary}

While mean-field models of heavy nuclei are useful for a qualitative understanding of deformation, they break the rotational invariance of the underlying Hamiltonian and exhibit a nonphysical sharp shape transition as a function of temperature. We have described a method, based on Eqs.~(\ref{prob-HS}) and (\ref{prob1}), to extract signatures of deformation in a framework that preserves rotational invariance. Qualitatively and at low temperatures, the mean-field characterization of deformed nuclei is supported by the new method. However, contrary to the mean-field approximation, the new method produces smooth shape transitions as a function of temperature. 

In the mean-field context, deformation is associated with an intrinsic frame. In the rotationally invariant framework, there is no well-defined intrinsic frame, since $\langle \hat Q_{2\mu} \rangle = 0$. However, $\hat Q_{20}$ can have a nontrivial distribution in the lab frame, which we have seen can provide a model-independent signature of deformation.

The quadrupole invariants have the same values in the lab frame and in the intrinsic frame. These values can in turn be expressed in terms of moments of the axial quadrupole in the lab frame, and this enables us to obtain information about the effective intrinsic deformation parameters in the rotationally invariant framework of the CI shell model, without resorting to a mean-field approximation. We have computed from AFMC these effective quadrupole shape parameters $\beta,\gamma$ for the isotope chains of even-mass  $^{148-154}$Sm and $^{144-152}$Nd nuclei.

For deformed nuclei, we have seen that the AFMC expectation $\langle \hat Q \cdot \hat Q \rangle$ is quite close to the HFB value at low temperature.

Nuclear deformation plays a key role in fission processes, where the level density as a function of deformation is a crucial input to models. Since the quadrupole projection method in the lab-frame allows us to extract information about the intrinsic shape through the use of quadrupole invariants, we will show in subsequent work that it can be used to calculate the level density as a function of excitation energy and intrinsic deformation parameters $\beta, \gamma$.

\section*{Acknowledgments} 

We thank J. Dobaczewski and K. Heyde for comments on the manuscript. This work was supported in part by the U.S. DOE grant Nos.~DE-FG02-91ER40608 and DE-FG02-00ER41132.
The research presented here used resources of the National Energy Research Scientific Computing Center, which is supported by the Office of Science of the U.S. Department of Energy under Contract No.~DE-AC02-05CH11231.  This work was also supported by the HPC facilities operated by, and the staff of, the Yale Center for Research Computing.

\section*{Appendix A: General solution for rotation angles} \label{appdx_angles}

In Sec.~\ref{sixangles} we derived a set of six angles which restore rotational invariance up to the second-order moment of $\hat Q_{20}$. Here we determine the angles that restore invariance up to the $n$-th order moment of $\hat{Q}_{2 0}$. (Note the six-angle solution is a special case; the general solution given here yields a set of ten angles for $n=2$.)

To begin, we express  $\hat Q^n_{20}$ as a sum over quadrupole invariants by successive coupling of angular momenta $2$ (the rank of $\hat Q$)
\begin{subequations} \label{q20n}
\begin{align}
    \hat{Q}_{20}^2 & =  \sum_J \cg{2}{0}{2}{0}{J}{0} (\hat Q \times \hat Q)^{(J)}_0 \;, \\
    \hat{Q}_{20}^3 & =  \sum_{J_{12},J} \cg{2}{0}{2}{0}{J_{12}}{0} \cg{J_{12}}{0}{2}{0}{J}{0}  \nonumber \\
    &  \qquad \qquad \times [(\hat Q \times \hat Q)^{(J_{12})} \times \hat Q]^{(J)}_0 \;, \\
    & \vdots \nonumber \\
    \hat{Q}_{20}^n & =  \sum_{\alpha} C_\alpha \, \hat{Q}^{(J)}_{\alpha,M=0} \;, 
\end{align}
\end{subequations}
where $\alpha=(J_{12}, J_{(12)3}, \ldots )$ labels the intermediate angular momenta and $\hat{Q}^{(J)}_{\alpha,M} = [[(\hat Q \times \hat Q)^{(J_{12})} \times \hat Q]^{(J_{(12)3})} \times \cdots  \hat \times \hat Q]^{(J)}_M$. In Eq.~(\ref{q20n}) $J$ takes on only even values $J=0,2,\ldots,2n$ because of selection rules.

Applying a rotation to $\hat{Q}_{20}^n$ we obtain
\be
\hat{R}(\Omega) \hat{Q}_{20}^n \hat{R}^\dagger(\Omega) = \sum_{M} D^{(J)}_{M,0}(\Omega) \sum_\alpha C_\alpha \, \hat{Q}_{\alpha,M}^{(J)} \;.
\ee
The general condition in order to zero out the terms for $2 \leq J \leq 2n$ by averaging over rotations $\hat{R}(\Omega_i)$ (such that only the scalar $J=0$ term survives) is then
\be
\hspace{-0.7em} \sum_{i=1}^{N_\Omega} D^{(J)}_{M,0}(\Omega_i) = 0, \;\; J=2,4,\ldots,2n,\; -J \leq M \leq J \,.
\ee

Averaging over these angles will then effectively restore rotational invariance sample-by-sample for the observables $\langle \hat{Q}_{20} \rangle, \langle \hat{Q}_{20}^2 \rangle, \ldots, \langle \hat{Q}_{20}^n \rangle$, and also reduce fluctuations in the distribution $P_\beta(q)$.

To find the angles, note that
\begin{align}
  D^{(J)}_{M 0}(\phi,\theta,\psi) & = \sqrt{\frac{4\pi}{2J+1}} Y_{J}^M(\theta,\phi) \nonumber \\
  & = \sqrt{\frac{(J-M)!}{(J+M)!}} P_J^M(\cos \theta) e^{i M \phi}\;.
\end{align}
Since $J \leq 2n$, we can project onto $M=0$ using a Fourier sum with angles $\varphi_k = 2\pi k /(2n+1)$, $k=0,1,\ldots,2n$:
\be
\sum_{k=0}^{2n} e^{-i M \varphi_k} = (2n+1) \delta_{M,0},\quad -2n \leq M \leq 2n \;.
\ee
Then
\be
\sum_{k=0}^{2n} D^{(J)}_M(\varphi_k,\theta) = (2n+1) \delta_{M,0} P_{J}(\cos \theta)\;.
\ee
We now need to find angles $\theta_l$ such that $\sum_l P_J(\cos \theta_l) = 0$. Or, defining $u_l \equiv \cos \theta_l$,
\be \label{PJzero}
\sum_l P_J(u_l) =0,\qquad J=2,4,\ldots,2n \;,
\ee
which is a set of $n$ independent equations and can be satisfied by a set of $n$ angles $\theta_l, l=1,2,\ldots,n$.

To solve, we express the even-powered monomials $u^{2m}$ in terms of $P_J(u)$,
\begin{align}
  u^{2m} & = \sum_{J=0,2,\ldots,2n} C_J^{(m)} P_J(u)\,,\\
  C_{J}^{(m)} & \equiv \frac{(2J+1)(2m)!}{2^{(2m-J)/2} ((2m-J)/2)! (J+2m+1)!!}\,,
\end{align}
for $m=0,1,\ldots,n$. Then using~(\ref{PJzero})
\be
  \sum_{l=1}^{n} u_l^{2m} 
    = C_{J=0}^{(m)} \sum_l P_0(u_l)  = n/(2m+1) \;.
\ee
So an equivalent set of equations to solve is
\be
\frac{1}{n} \sum_{l=1}^n u_l^{2m} = \frac{1}{2m+1},\qquad m=1,2,\ldots,n \;.
\ee
The set of $n(2n+1)$ angles $(\theta_l,\varphi_k)$ then satisfy
\be
\frac{1}{n(2n+1)} \sum_{l=1}^n \sum_{k=0}^{2n} D^{(J)}_{M 0}(\theta_l,\phi_k) = \delta_{M,0} \delta_{J,0}
\ee
for $J=0,2,\ldots,2n$ and $-2n \leq M \leq 2n$.

\subsubsection{10-angle solution for $n=2$}

Setting $n=2$ in the general solution gives a set of 10 angles, different from the 6 angles discussed in Sec.~\ref{sixangles}. The azimuthal angles are
\be
\varphi_k = {2 \pi k/5}, \qquad k=0,1,\ldots,4 \;,
\ee
and the angles $\theta_l$ are determined by the equations
\begin{align}
  u_1^2 + u_2^2 &=  2/3 \\
  u_1^4 + u_2^4 &=  2/5 \;.
\end{align}
with solution
\begin{align}
  \cos^2 \theta_1 & = (1+2/\sqrt{5})/3 \\
  \cos^2 \theta_2 & = (1-2/\sqrt{5})/3 \;.
\end{align}

\subsubsection{21-angle solution for $n=3$}
For $n=3$, the azimuthal angles are
\be
\varphi_k = {2 \pi k/7}, \qquad k=0,1,\ldots,6 \;,
\ee
and the $\theta_l$ are determined by the equations
\begin{align}
  u_1^2 + u_2^2 + u_3^2 &= 1 \\
  u_1^4 + u_2^4 + u_3^4 &= 3/5 \\
  u_1^6 + u_2^6 + u_3^6 & = 3/7 \;,
\end{align}
with solution
\begin{align}
  \cos^2 \theta_1 &= 0.7504\ldots \\
  \cos^2 \theta_2 &= 0.1785\ldots \\
  \cos^2 \theta_3 &= 0.0711\ldots  \;.
\end{align}
This gives a set of 21 angles that is sufficient to thermalize the distribution of $\hat Q_{20}$, as shown in Fig.~\ref{sm_zvq_avging}.

\section*{Appendix B: Lab-frame quadrupole moments for the rigid rotor}

The moments of the rigid-rotor distribution (\ref{prob-rotor}) can be calculated in terms of $q_0$ from a simple recursion relation. To calculate the moments $\langle q^n\rangle$, we derive a recursion formula as follows. Measuring $q$ in units of $q_0$, i.e., $x=q/q_0$,  we use integration by parts to find
\begin{align}
  \int \!\! d x {x^n \over \sqrt{1+2x}}  & = x^n \sqrt{1+2x} -n \!\! \int \!\! d x \, x^{n-1}\sqrt{2x+1}  \nonumber \\
  & = x^n \sqrt{1+2x} -n \!\! \int \!\! d x \, x^{n-1}{2x+1\over\sqrt{2x+1}} \nonumber
\end{align}
or
\be
(2n+1) \!\! \int \! d x {x^n \over \sqrt{1+2x}} = x^n \sqrt{1+2x} -n \!\! \int \! d x { x^{n-1} \over \sqrt{2x+1}}\;.
\ee
Taking the limits between $-1/2$ and $1$, the $n$-th moment $\langle x^n\rangle= \frac{1}{\sqrt{3}}\int_{-1/2}^1 d x \, {x^n \over \sqrt{1+2x}}$ satisfies the recursion relation
\be\label{moment-recursion}
\langle x^n\rangle = {1 \over 2n+1} \left(1-n\langle x^{n-1}\rangle\right) \;.
\ee
Starting with $\langle q^0\rangle=1$ (normalization), we can use Eq.~(\ref{moment-recursion}) to calculate the first few moments
\be
\langle q \rangle =0\;;\; \langle q^2 \rangle = \frac{1}{5} q_0^2\;;\; \langle q^3 \rangle = \frac{2}{35} q_0^3
\;;\; \langle q^4 \rangle = \frac{3}{35} q_0^4 \;;\ldots \;.
\ee

\end{document}